# Runtime Adaptability driven by Negotiable Quality Requirements

Adina Mosincat

University of Lugano, Switzerland
`adina.diana.mosincat@usi.ch`

**Abstract.** Two of the common features of business and the web are diversity and dynamism. Diversity results in users having different preferences for the quality requirements of a system. Diversity also makes possible alternative implementations for functional requirements, called variants, each of them providing different quality. The quality provided by the system may vary due to different variant components and changes in the environment. The challenge is to dynamically adapt to quality variations and to find the variant that best fulfills the multi-criteria quality requirements driven by user preferences and current runtime conditions. For service-oriented systems this challenge is augmented by their distributed nature and lack of control over the constituent services and their provided quality of service (QoS). We propose a novel approach to runtime adaptability that detects QoS changes, updates the system model with runtime information, and uses the model to select the variant to execute at runtime. We introduce negotiable maintenance goals to express user quality preferences in the requirements model and automatically interpret them quantitatively for system execution. Our lightweight selection strategy selects the variant that best fulfills the user required multi-criteria QoS based on updated QoS values.

**Keywords:** runtime adaptability, quality requirements, QoS , user preferences, service composition

## 1 Introduction

Systems are increasingly required to run in a dynamic environment where runtime adaptability is crucial. The diversity makes possible alternative implementations for functional requirements, called variants. Dynamically adaptive systems ensure functional requirements are met by switching between variants at runtime.

Quality requirements are important aspects of a software system. There are two important issues to be taken into consideration when quality requirements are concerned: firstly, users may have different, sometimes conflicting, preferences regarding quality requirements, and secondly, the delivered quality estimated at design time may vary due to parameter changes at runtime. The system not only has to meet the users' different quality expectations, but it also has to adapt to changes in the environment to meet these requirements. Thus, runtime adaptability is also crucial when quality requirements of the system are concerned.

Service compositions allow reusing existing functionalities to build new applications. The quality requirements of a service oriented system are expressed as Quality of Service (QoS) parameters guarantees, called service level objectives (SLOs [1]), in service level agreements (SLAs [1]). Usually, only one exclusive SLO for each QoS parameter is provided for all users. However, in the diverse and dynamic web environment, it is vital that the system is aware of different user preferences and is able to meet different user expectations. We address this issue by introducing the negotiable maintenance goals to allow users to express their QoS preferences in the requirements model. The negotiable maintenance goals allow for flexibility but do not leave room to misinterpretations when the user preferences are interpreted quantitatively.

We address the issue of runtime adaptability by using a model annotated with QoS values. The model is updated with runtime information and the QoS values are re-computed when changes occur. The model is used at runtime to select the variant execution that offers the delivered quality closer to the user preferences. The selection strategy takes into consideration the user preferences expressed by the negotiable maintenance goals and the updated QoS values that reflect the current runtime conditions.

In previous work [29] we have introduced an approach to runtime adaptability making use of an evolving model at runtime to ensure the unique SLO of the system is met. The work presented in this paper focuses on handling different user preferences that are expressed at the model level and must be met at runtime, thus allowing for multiple SLOs.

The contributions of this paper are two-fold: (1) We introduce a novel approach to express user preferences regarding quality requirements in the requirements model and ensure their fulfillment down to the system execution level. We present a taxonomy of negotiable multi-criteria QoS that are automatically converted into mathematical formulas evaluated at runtime. (2) We present a lightweight strategy for optimal runtime variant selection. Together with the change detection mechanism, our strategy allows the system to adapt to changes that affect the provided quality.

This paper is structured as follows: Section 2 introduces our approach. Section 3 details the system models and the negotiable maintenance goals, while the variant selection strategy is described in Section 4. We present evaluation results in Section 5, discuss related work in Section 6, and conclude with Section 7.

## 2 Approach Overview

Fig. 1 shows the architecture of our approach. The grey boxes represent third-party components, and the dashed boxes group the components according to four main activities in our approach - express, find, (re)estimate, execute:

1. Quality requirements are expressed as negotiable maintenance goals attached to functional requirements in the *Requirements Model*.

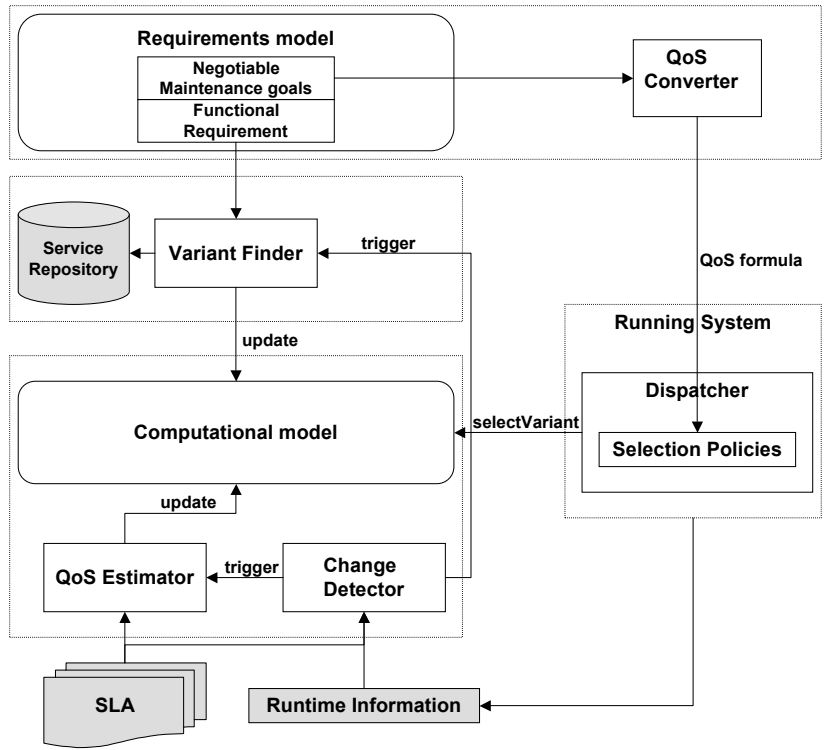

**Fig. 1.** Approach Overview.

2. The *Variant Finder* automatically finds variants, i.e. implementations, for each functional requirement, generating the computational model.
3. Variants from the *Computational Model* are annotated with quality values that are (re-)estimated on change based on runtime information.
4. Variants selected by the Dispatcher of the *Running System* are executed.

**Express**

Users have different preferences with regard to quality requirements. Often users are willing to negotiate the quality requirements according to what the system can offer and to the importance of each QoS parameter for them. The system must be aware of the different user preferences, which implies that the user must be able to express his preferences, and the system must be able to interpret these preferences.

In the GORE technique, *maintenance goals* are used to express the system quality requirements. Usually, maintenance goals are global, i.e., one for all users, and leave room to misinterpretations when the requirement is quantified. We introduce *negotiable maintenance goals* to allow users to express specific preferences. We define a set of macros that are used by the QoS Converter to convert the negotiable maintenance goals into mathematical formulas which are implemented by Selection Policies and evaluated during the variant selection.

**Find**

The diversity of the business landscape and the burst of the web services makes possible that a functional requirement of a system can be implemented in different ways. Variants implementing the same functional requirement may use completely different components and provide different quality. In this paper we strictly refer to variants expressed as service compositions.

The Variant Finder is responsible with matching functional requirements with existing variants, i.e. implementations. The Variant Finder can be a human, and thus the variants are manually provided, or it can be an automated system such as automatic composer systems ([2]) that query the service repository finding the set of service compositions which solve the requirement. The automatic composer systems impose constraints such as expressing the functional requirement in a query language the composer understands.

Usually service discovery and matchmaking is a time consuming process. A set of variants solving the functional requirement are provided statically at design time. The search for a new variant can be triggered upon receiving runtime information signaling a significant change, such as the availability of a new service. The Variant Finder updates the computational model when it finds a new variant for the functional requirement.

**(Re)Estimate**

The initial QoS values are computed for each variant based on parameters available at design time. However, given the dynamic nature of services, these parameters may change during runtime and affect the system provided quality. The Change Detector signals when a significant change occurs and triggers the re-estimation of the provided quality. The QoS Estimator re-computes the QoS parameter values for each affected variant based on runtime information and updates the computational model with the new values.

**Execute**

Some systems that are an aggregation of distributed applications, such as service-oriented systems, can be seen as a composition of functional requirements implemented independently. Each variant is a service composition. The system core is a dispatcher that forwards requests to the variant selected for execution. The variant is selected according to selection policies that implement the user QoS preferences and according to QoS values. Monitoring information is gathered during the variant execution and is used along with third-party runtime information in the re-estimate activity.

Activities that are time-consuming and computational intense, namely express, find and initial estimation of QoS values, are performed statically when the computational model is generated.

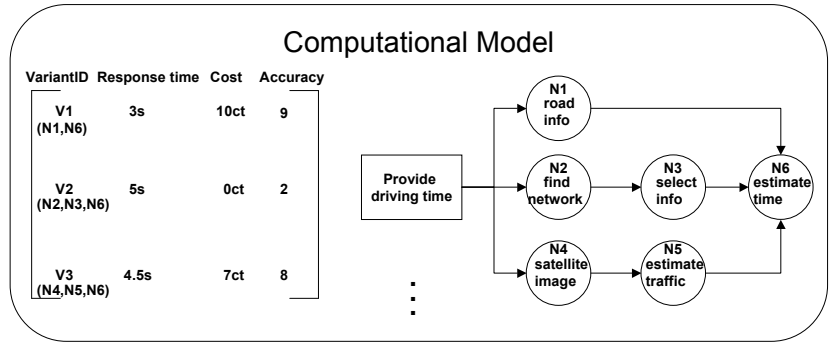

**Fig. 2.** Part of the computational model detailing the "provide driving time" functional requirement. Response time expressed in seconds, cost in cents.

## 3 System Models

We use as requirements model a goal model with functional goals (functional requirements) as presented in [3] and attach negotiable maintenance goals to the functional goals. We generate the computational model from the requirements model by providing variants for the functional requirements.

Fig. 2 shows a fragment of the computational model for an intelligent route-planner system. The functional requirement detailed in the figure is "provide the time needed to drive between two locations based on traffic conditions". Each node in the graph represents a service. The Variant Finder provides three variants that solve the functional requirement:

1. The system gets traffic information for the route area from the road department (node N1 road info in the figure) and estimates the driving time considering the available information such as trip length, speed limitations, and traffic queues (Node N6 estimate time).
2. The system searches drivers networks and gathers information about the traffic in the route area (node N2 find network), it filters the gathered information (node N3 select info) and estimates the driving time (node N6).
3. The system gets road images for the planned route from a satellite database (node N4 satellite image), and computes traffic estimates by comparing the retrieved images (node N5 estimate traffic). It then computes the driving time based on the traffic estimates (node N6).

The functional requirement in the computational model is annotated with the QoS values computed for each variant. Three QoS parameters are considered: response time, cost, and the user defined parameter accuracy. The parameter accuracy of the provided time estimation has values in the integer interval [0–10]. At model generation, the QoS values are computed based on information from the providers of the composing services, e.g. available in the services' SLAs.

### 3.1 Negotiable Quality Requirements

Often users have different preferences regarding the system requirements. Quality requirements expressed in natural language leave room to misinterpretations. For instance, a user might say that cost is the only parameter that matters, but he might change his mind when faced with low values of other parameters such as response time. The negotiable maintenance goals allow for flexibility and avoid misinterpretations by means of thresholds and priorities. Thresholds allow users to specify the tolerance limits for QoS parameter values. Priorities are weights assigned to each QoS parameter reflecting the importance of the parameter to the user. We use the following taxonomy of negotiable maintenance goals:

1. *High Priority* – the user can define one QoS parameter that has high priority. This means that the system should try to achieve the best value for the specified QoS parameter, regardless of the values of the other QoS parameters. The user can optionally fix thresholds for all parameters, i.e. the worst value he is willing to accept for the parameters. Fixing thresholds applies to all types of negotiable maintenance goals.
   E.g.: Response time is high priority. Response time is less than 1 day. Cost is less than 10 euros.
2. *Distributed Priority* – the user can choose several QoS parameters that have the same level of priority, while all other parameters are ignored.
   E.g.: Response time, cost is high priority. Maximum time to recovery is less than 1 day.
3. *Conditional Priority* – the user chooses one or more QoS parameters that can vary (degrade) if one or more parameters with higher priority varies (upgrades).
   E.g.: If cost upgrades by 20% then response time degrades by 20%.

We define the following macros that map the negotiable maintenance goals to mathematical formulas that are implemented in the Selection Policies logic which we use at runtime for the variant selection:

1. $list(parameters)$ IS_HIGH_PRIORITY maps to
   $priority_{P_i} = 1/m, \forall P_i \in list(parameters)$
   $priority_{P_j} = 0, \forall P_j \notin list(parameters)$,
   where m is the number of QoS parameters in the list, and $P$ stands for parameter.
2. $list(parameters)$ IS_LESS_THAN $value$ maps to $threshold_{P_i} = value$, $\forall P_i \in list(parameters)$. Used for negative parameters, such as response time (lower values are better).
3. $list(parameters)$ IS_GREATER_THAN $value$ maps to $threshold_{P_i} = value$, $\forall P_i \in list(parameters)$. Used for positive parameters, such as accuracy (higher values are better).
4. IF $list(firstparam)$ UPGRADES_BY $firstvalue$ THEN $list(secondparam)$ DEGRADES_BY $secondvalue$
   The $firstvalue$, respectively $secondvalue$ are percentages. The macro is mapped to the following equations:

(a) For negative parameters.
$threshold_{P_i} = threshold_{P_i} + (firstvalue * threshold_{P_i})/100$
$threshold_{P_j} = threshold_{P_j} - (secondvalue * threshold_{P_j})/100$,
$\forall P_i \in list(firstparam), \forall P_j \in list(secondparam), P_j \neq P_i$.

(b) For positive parameters.
$threshold_{P_i} = threshold_{P_i} - (firstvalue * threshold_{P_i})/100$
$threshold_{P_j} = threshold_{P_j} + (secondvalue * threshold_{P_j})/100$,
$\forall P_i \in list(firstparam), \forall P_j \in list(secondparam), P_j \neq P_i$.

(c) For both positive and negative parameters, the priorities are modified according to the following pseudocode, where $P_i \in list(firstparam)$, $P_j \in list(secondparam)$, $list(firstparam) \cap list(secondparam) = \emptyset$, $m$ is the number of parameters in the $firstparam$ list, and $n$ is the number of parameters in the $secondparam$ list:

$max = maximum(firstvalue, secondvalue)$;
if $max > 100$ then begin
   $priority_{P_i} = 1/m$;
   $priority_{P_j} = 0$;
end
else begin
   $min = minimum(priority_{P_i})$
   $difference_j = priority_{P_j} - min * max$;
   $priority_{P_j} = min * max$;
   $priority_{P_i} = priority_{P_i} + \sum_{j=1}^{n} difference_j/m$;
end

In the example in Fig. 2, there are at least two different user preferences regarding the QoS:

1. Users for whom time is vital need to get the information as soon as possible, for instance to be able to find alternative routes in case of traffic jams. These users agree to pay up to a certain amount for the information. The corresponding negotiable maintenance goal is: Response time is high priority. Response time is less than 4s. Cost is less than 10 ct.
2. Users who are not under time pressure are interested in the information, but only if it comes for free. These users agree to wait to get the information. The corresponding negotiable maintenance goal is: Cost is less than 0 ct.

## 4 QoS Estimation, Variant Selection and Change Detection

The cumulative end-to-end QoS of a service composition depends on the individual QoS of each constituent service. The modification of the QoS parameters of one service can therefor affect the QoS of the variants that use the service. Various techniques have been proposed to tackle the optimization of the cumulative QoS at the level of service selection, resulting in different mathematical optimization models, such as multiple attribute utility theory and regression models [4,5]. Usually these are time and resource consuming solutions.

Although our approach can integrate these solutions, for performance as well as visibility considerations we use a fixed set of variants provided at model generation time and preserve the constituent services of the variant. We compute the cumulative QoS parameters of each variant based on the formulas introduced in [6]. The initial QoS computation formulas derived at model generation time are stored in the system and used for QoS re-computation. When a QoS parameter of a service changes, the QoS parameter is re-computed for all variants using the service.

When selecting the variant to execute the Dispatcher takes into consideration the user preferences as expressed by the negotiable maintenance goals and stored in the Selection Policies. The negotiable maintenance goals set the *threshold* and *priorities* values which are used by the variant selection strategy.

The selection strategy ensures that the variant that best fulfills the user preferences is selected from the available variants. The strategy uses the latest updated values of the QoS parameters computed for each variant based on runtime information.

We consider $m$ variants and $n$ QoS parameters. For all parameters for which no threshold values are given

$$threshold_{P_i} = max(P_i monitored),$$

where $threshold_{P_i}$ is the threshold value for parameter $P_i$ and $max(P_i monitored)$ is the maximum monitored value for parameter $P_i$.

For each variant we compute the deviation from the threshold in percentage for each parameter as

$$\sigma_{P_{ij}} = (P_i crt - threshold_{P_i}) * 100/threshold_{P_i}, 1 \leq i \leq n, 1 \leq j \leq m,$$

where $P_i crt$ is the last computed value for parameter $P_i$. All variants for which $\exists \sigma_{P_{ij}} > 0, \forall i,j$ where $1 \leq i \leq n, 1 \leq j \leq m$ and $P_i$ is a negative parameter, respectively $\exists \sigma_{P_{ij}} < 0, \forall i,j$ where $1 \leq i \leq n, 1 \leq j \leq m$ and $P_i$ is a positive parameter are excluded from selection. This means that all variants that have at least one parameter value exceeding, respectively not reaching, the user defined threshold are excluded from selection.

We choose the variant $v$, $1 \leq v \leq m$ with

$$\sum_{i=1}^{n} priority_{P_i u} * \sigma_{iv} = \min \sum_{i=1,j=1}^{n,m} priority_{P_i u} * \sigma_{P_{ij}},$$

where $priority_{P_i u}$ is the priority of parameter $P_i$ for user $u$, and

$$\sum_{i=1}^{n} priority_{P_i u} = 1.$$

Change Detectors can determine when a critical change has occurred and trigger the re-estimate QoS activity. An example of a change that could have a high impact on the QoS of a variant is a violation of the SLOs of one of the constituent services. Assuming the service SLA is available, a dedicated change detector can use monitoring information and implement tests to determine if a violation has occurred by using statistic methods, such as statistical hypothesis testing [7].

In order for statistical tests to be effective, a large number of samples, i.e. monitoring values, might be needed. Since the system relies upon monitoring and runtime information provided by third parties, a simpler Change Detector testing for fluctuations in the QoS parameter values might be better suited.

We explain below the Change Detector we use in our evaluation. The Change Detector detects service QoS fluctuations by computing the deviation of the last received QoS parameter value from the average QoS parameter value. The average QoS parameter value is stored in the system and computed from all previous values received. If a significant QoS parameter fluctuation is detected, the Change Detector triggers the re-computation of the QoS parameter values of all variants using the service. The logic implemented by the Change Detector is expressed below. $monitoredQoS_i{}^{sid}$ is the monitored QoS parameter value of the service identified by sid and $referenceQoS_i{}^{sid}$ is the QoS parameter average value of service sid stored in the system:

**if** $\left|monitoredQoS_i{}^{sid} - referenceQoS_i{}^{sid}\right| > threshold$ **then**
    trigger()
**end**

The configurable parameter threshold controls the significance of the QoS parameter fluctuation. For QoS parameters that are not expected to fluctuate, such as cost, the threshold is 0. For QoS parameters that are usually suffering moderate fluctuations, such as response time, the value of the threshold parameter is computed as function of the average value, e.g. $1/2 * referenceQoS_i{}^{sid}$.

## 5 Evaluation

In this section we evaluate our approach by exploring the ability of the system to meet the user preferences considering the availability of variants with different QoS values, respectively in the presence of changes that affect the provided variant QoS. We assess the effectiveness of our selection strategy in settings with high numbers of variants, QoS parameters and user requests.

We base our evaluation on the intelligent route-planner example. For the implementation of the driving time functional requirement, our system provides 5 variants each of them implemented as BPEL processes. As source of runtime information in terms of response time we use ADULA [8,9], a framework for monitoring BPEL processes. We simulate a third-party runtime information provider for the cost and accuracy QoS parameters.

We use two different settings to explore the correctness of our selection strategy for three different user negotiable maintenance goals, high priority, distributed priority, respectively conditional priority: (1) In the first setting we show how the variant selection strategy works when the environment is stable, i.e., no changes occur. (2) In the second setting we explore the system behavior in the presence of response time fluctuations and cost changes. We show how the system adapts to changes and how the changes affect the variant selection.

**Table 1.** Left side: Initial values for QoS parameters for each variant.
Right side: Threshold ($T_{RT}$, $T_C$, $T_A$), priority ($P_{RT}$, $P_C$, $P_A$) values, and selected variant (Sel) for high priority, distributed priority and conditional priority goals.

| QoS parameter | V1 | V2 | V3 | V4 | V5 | Goal | $T_{RT}$ | $P_{RT}$ | $T_C$ | $P_C$ | $T_A$ | $P_A$ | Sel |
|---|---|---|---|---|---|---|---|---|---|---|---|---|---|
| Response time (RT) | 3s | 5s | 4.5s | 10s | 3.5s | High | 10s | 1 | 20ct | 0 | 2 | 0 | V1 |
| Cost (C) | 10ct | 0ct | 7ct | 20ct | 8ct | Distributed | 10s | 1/3 | 20ct | 1/3 | 5 | 1/3 | V5 |
| Accuracy (A) | 9 | 2 | 8 | 10 | 9 | Conditional | 4.8s | 1/10 | 8ct | 9/10 | 2 | 0 | V3 |

We use 10 different services for the 5 variants implementing the driving time functional requirement. The service that estimates the driving time based on the road information is used by four of the variants (V1 – V4) and we identify it as service S, as we will use it further for describing the evaluation settings.

Variant V1 uses two services: one service that provides road information in terms of number of cars on the road sector and average driving speed, and service S. Variant V2 uses 3 different services: one service to gather road information from the network, one service to select meaningful information from the gathered road information, and service S. Variant V3 uses 3 different services: one service to retrieve a satellite image of the road sector, one service to compute road information from the satellite image and service S. Variant V4 uses 4 different services: one service to retrieve several images of the targeted area taken from helicopters surveilling the area, one service to compare the images and render one composed image of the road sector, one service to compute road information from the composed image and service S. Variant V5 uses three services, one service to read road sensors information, one service to compute the number of cars and average driving time based on the road sensors information, and one service to estimate the driving time based on the road information and weather conditions.

We consider the following three negotiable maintenance goals, high priority, distributed priority, respectively conditional priority:

1. Response time is high priority.
2. Response time, cost, accuracy is high priority. Accuracy is greater than 5.
3. If cost upgrades by 20% then response time degrades by 20%. Cost is less than 10 ct. Response time is less than 4 s.

Left side of Table 1 shows the QoS values computed for each variant when the computational model is created. Right side of Table 1 shows the thresholds, priorities and the selected variant for each of the three negotiable maintenance goals. For the high priority goal, only parameter response time is considered for the variant selection. Variant $V1$ is selected because it has the best response time value.

For the distributed priority goal all three QoS parameters influence the selection. Variant $V2$ with accuracy below the user defined threshold is excluded from selection. Variant $V5$ with the best overall QoS is selected.

For the conditional priority goal the user accepts a degradation of the response time parameter if the cost is lower. When the cost is lower than 8 ct (20% deviation from the cost threshold), the response time could be higher than 4 s. However, the response time should not be higher than 4.8 s (20% deviation from the threshold). Variant $V3$ is selected as it offers the best QoS values meeting these requirements.

**Table 2.** Selected variant with updated QoS parameter values computed after each change. Changes are: (1) service S slows down by 0.6s, (2) service S slows down by 0.9s, (3) service S returns to initial state (speeds up by 1.5s), and cost is decreased, (4) service S slows down by 0.6s, (5) service S slows down by 0.9s. $RT_S$ is the response time of service S, $C$ is the cost of service used by V1.

| Goal | $RT_S = 1.6s$ $C = 10ct$ | $RT_S = 2.5s$ $C = 10ct$ | $RT_S = 1s$ $C = 6ct$ | $RT_S = 1.6s$ $C = 6ct$ | $RT_S = 2.5s$ $C = 6ct$ |
|---|---|---|---|---|---|
| **High** | V5{3.5s,8ct,9} | V5{3.5s,8ct,9} | V1{3s,6ct,9} | V5{3.5s,8ct,9} | V5{3.5s,8ct,9} |
| **Distributed** | V5{3.5s,8ct,9} | V5{3.5s,8ct,9} | V1{3s,6ct,9} | V1{3.6s,6ct,9} | V5{3.5s,8ct,9} |
| **Conditional** | V5{3.5s,8ct,9} | V5{3.5s,8ct,9} | V1{3s,6ct,9} | V1{3.6s,6ct,9} | V1{4.5s,6ct,9} |

In our second evaluation setting, firstly we vary the response time of service S which influences the response time of variants $V1$, $V2$, $V3$ and $V4$. For service S we use a discrete time Markov chain [24] service model with 3 different states. In the first state the response time of service S is 1s, in the second state the response time of S is 1.6s, and in the third state the response time of S is 2.5s. The service changes state every 5 requests. We use the Change Detector to detect when the response time changes. The Change Detector uses the monitoring information provided by ADULA. Secondly, after the first 15 requests, we change the cost parameter of the road information service used by variant $V1$ by 40%.

Table 2 shows the selected variant after every change with the new computed QoS values in curly brackets. As service S is used by variants $V1$–$V4$ the response time value of these variants is re-computed when the response time of service S changes. Variant $V5$ remains unchanged. The updates of the response time values result in different variant selections for the three negotiable maintenance goals. Thus, when service S becomes slow, variant $V5$ becomes the fastest available variant and is selected for all goals.

We change the cost of the service used by variant $V1$ so that it decreases from 10ct to 6ct. Since this is the only service with cost used by variant $V1$, the cost value for variant $V1$ is re-computed and becomes 6ct. Given the updated cost value, variant $V1$ is selected for the distributed and conditional priority goal.

In the third setting we assess the effectiveness of our selection strategy for complex systems. We use an abstract functional requirement for which we consider 200 variants and 20 QoS parameters and measure the time needed for variant selection. All variants use the sequential workflow pattern. As expected, the time needed for variant selection is negligible, up to 2 ms per user request, respectively an average of 280ms to select the variant for 300 user requests (the requests are started by 5 concurrent threads).

In conclusion, our selection strategy chooses the best variant available to fulfill the user different quality expectations. When changes occur, the system adapts to these changes switching between existing variants to preserve the provided quality that best meets the user different quality expectations.

## 6 Related Work

There are different approaches that address finding the optimal solution to meet the cumulative multi-criteria QoS of a service composition, i.e. variant. Some of these solutions analyze and compose the QoS of individual services to obtain the QoS of the process statically at design time [10,11]. In contrast, our solution is dynamic and we consider changes in the environment. Other solutions continuously recompute the estimated QoS [4,5,12,13,14]. These approaches propose a mathematical model based on which the constituent services are selected at runtime. In contrast to our approach, these solutions can also adapt to changes that occur during the variant execution. On the other hand, they are computational expensive. While our approach can integrate these solutions, for performance considerations we prefer to use a set of fixed variants that are provided at model generation. The QoS of the variants is re-computed on change and our selection strategy uses the updated values allowing for inexpensive variant selection.

A solution using machine learning techniques is presented in [5]. The solution uses regression models and re-computes the QoS parameter values at different points during the process execution, specified initially by the process developer. The solution allows for high accuracy of prediction and can react to changes that occur between the defined checkpoints, but it has to pay the price in computational expense and to rely on the developer's ability of instrumenting the process. In contrast, our solution requires user awareness in the requirement analysis phase and offers complete transparency to the developer.

In [14] the authors reason about the possible outcomes of the overall quality of a composition when a service is selected based on Graph Transformation System. The method is used to predict all possible architectural configurations that can be reached by selecting a service.

A flexible approach allowing for integration of user preferences in the computation of QoS parameters that affect service selection is the LCP-nets framework [4]. When computing the QoS of the process the framework takes into consideration user preferences, relative importance, and tradeoffs between QoS parameters, which are expressed in linguistic terms using LCP-nets. It then compares the candidate services upon several different QoS dimensions, applying the expressed preferences to the currently measured QoS values. As our solution does not consider selection of individual composing services, it allows for less computational expensive variant selection.

A different direction is taken by the Planning as Model Checking [15] approach, which allows to monitor the execution of the composition and take corrective actions in case some of the conditions are not met. The approach uses planning techniques as well as the EaGLe goal language [16], a language that allows expressing system goals such as non-functional requirements.

Another approach to deal with unsatisfactory QoS provided by services is SLA negotiation [17,18,19,20]. The negotiation can be manual, requiring human intervention, or automatic in which case software agents are used. Negotiation of SLA is done on a client basis, which means that a service can have a different SLA for each of its consumers. These solutions are based on agents that carry out negotiation, i.e., explore possible solutions that eventually lead to an agreement, using different algorithms and negotiation strategy models. While some approaches use optimization algorithms to speed up the negotiation process [17], finding an agreement may be a long-lasting process.

[21] takes a new view on contract (SLA) negotiation, considering the evolution of the contract based on the possible evolution of the parties involved in the contract. The solution provides a way of defining constraints on the contract, defining boundaries in which the provided service QoS can vary and what is acceptable both to provider and consumer. Therefore, a contract violation is more strictly defined in an evolving context and the need of re-negotiation is reduced.

In [22] the authors address the need to allow for flexibility of quality requirements by defining constraints in the requirement model. The norms ensure the non-functional properties of the service are expressed in the requirements model and service warranties protect the stakeholder from variations of the norms. This approach stops at the requirements model level without considering an evolution of the model based on runtime information.

Recent focus on encapsulating adaptation at the requirements model level led to a number of interesting solutions. KAMI [23] is a framework that provides information on non-functional values based on current runtime conditions. KAMI uses DTMC [24] models to reason about non-functional properties of the system and bayesian techniques to estimate their values based on runtime information. Our solution also handles different user preferences and uses the model at runtime to allow the system to adapt to changes that affect the provided quality.

An interesting solution that makes use of a goal-oriented model at runtime is [25]. The authors introduce adaptive goals to allow specification of different adaptive features in the requirements model and trigger adaptation actions when the satisfaction level of a goal deviates from the required value. Our solution focuses on specifying user preferences in the model which then drive the system execution.

CCAP [26] is a system that provides support for configurable and adaptive service compositions aware of user context and different needs. The system makes use of a configurable composition model which provides abstractions for service context and exceptions, and of an execution model based on control tuples, favoring robust and adaptive service execution. Another approach to composition adaptability at the architectural level is taken in [27] which introduces Adaptive Service Oriented Architecture (ASOA). ASOA merges the autonomic computing with SOA. A solution allowing for self-reconfiguration to adapt and recover from failures is proposed in [28], which introduces a conceptual architecture to provide systems with adaptive capabilities.

[29] introduces an approach close to ours to ensure response time SLOs of the system are met regardless of the changes in the environment. The response time of the variants is monitored and stored in the system to detect based on statistic tests when an SLO violation occurs. The approach considers only the response time parameter and one exclusive SLO for all users of the system. Our approach focuses on different user preferences, which might result in multiple SLOs, and take into consideration multi-criteria QoS.

[30] presents a solution to provide the requirement engineer with feedback on service behavior gathered using online testing techniques. Based on the feedback the requirement engineer can decide on adaptation actions to execute. In contrast, our solution automatically adapts to changes that affect the QoS parameter values and does not require manual modification of the requirements model.

ALBERT [31] is a language that is used to specify functional and non-functional property assertions which are verified at design time by model checking and used at runtime as dynamically evaluated assertions. Our approach takes dynamic models into consideration, i.e., models that can evolve at runtime because of environmental changes.

Runtime adaptation in our approach leverages the idea of using software models at runtime introduced in Models@Run.Time [32].

# 7 Conclusion

In this paper we have presented a novel solution for runtime adaptability that enables the system to provide different quality in order to meet the user QoS preferences. We have introduced negotiable maintenance goals to allow expressing user preferences. Our approach avoids misinterpretations of quality requirements by automatically converting user preferences into mathematical formulas evaluated at runtime. The solution generates a computational model from the system requirements model and annotates it with QoS values. The QoS values are updated based on runtime information to reflect changes in the environment.

The choice of the variant executed at runtime is driven by the negotiable maintenance goals and uses the updated model. In this way our solution ensures that the system adapts to changes in the environment so as to provide the quality that best fulfills the users different expectations.

The solution depends on the availability of runtime information. We have evaluated our approach integrating an existing monitoring system for BPEL processes, and shown that our selection strategy selects the best available variant to fulfill different user quality requirements and that our system adapts to changes of parameters which affect the provided quality.